\documentstyle[aps,preprint]{revtex}
\begin{document}
\draft
\def\ao{\hat{a}}       %
\def\iao{\ao^{-1}}
\def\iaom{\ao^{-m}}
\def\coh{\vert \alpha \rangle}
\def\co{{\hat{a}^\dagger}}
\def\com{\hat{a}^{\dagger m}}
\def\cot{\hat{a}^{\dagger 2}}
\def\ico{\ao^{\dagger -1}}
\def\icom{\ao^{\dagger -m}}
\def\opfn{f(\hat{n})}
\def\opfnm{f(\hat{n},m)}
\def\ecoh{\vert\alpha,m\rangle}
\def\necoh{\vert\alpha,-m\rangle}
\def\ns{\vert n\rangle}

\bibliographystyle{unsrt}

\title{Photon-added coherent states as nonlinear coherent states } 

\author{S. Sivakumar \\
      Laboratory and Measurements Section\\ 307, General Services Building\\
	  Indira Gandhi Centre for Atomic Research\\
	  Kalpakkam, INDIA- 603 102.\\
       }
\maketitle

\begin{abstract}

	The states $\ecoh$, defined as $\com\coh$ up to a normalization 
constant and $m$ is a nonnegative integer, are shown to be the eigenstates of
  $\opfnm\ao$ where $\opfnm$
is a nonlinear function of the number operator $\hat{n}$.  The explicit 
form of $\opfnm$ is constructed.  The eigenstates of this operator for 
negative values of $m$ are introduced. The  properties of these
 states are discussed and compared with those of the state $\ecoh$. 
\\

\end{abstract}

\section{INTRODUCTION}

	Coherent states are important in many fields of physics\cite{jrk,rjg}.
  Coherent states $\coh$, defined as the eigenstates of the harmonic
oscillator annihilation operator $\ao$, $\ao \coh = \alpha \coh$ \cite{gla},
 have properties like the classical radiation field.  There exist states of the 
electromagnetic field whose properties, like squeezing, higher order
squeezing, antibunching and sub-Poissonian statistics \cite{dfw,loud}, are
 strictly quantum mechanical in nature. These states are called as nonclassical
 states. The coherent states define the limit between the classical and
 nonclassical behaviour of the radiation field as far as the nonclassical
 effects are considered.  A generalization of the coherent states was done by
 q-deforming the basic commutation relation $[\ao,\co] =1$\cite{qcs,mfl}. 
 A further generalization is to define states that are eigenstates of the
 operator $\opfn \ao$,
\begin{equation}
\opfn\ao\vert f,\alpha\rangle=\alpha\vert f,\alpha\rangle,
\end{equation}
 where 
$\opfn$ is a operator valued funtion of the number operator $\hat{n} = \co 
\ao$. These eigenstates are called as nonlinear coherent states and they are
nonclassical.  In the linear limit, $\opfn =1$, the nonlinear coherent states
 become the usual coherent states $\coh$.  The nonlinear coherent states were
 introduced, as f-coherent states, in connection with the study of the 
oscillator whose frequency depends on its energy\cite{manko}.  A class of
 nonlinear coherent states can be realized physically as the stationary states
 of the center-of-mass motion of a trapped ion\cite{filho}.  These nonlinear
 coherent states exhibit nonclassical features like squeezing and 
 self-splitting.  

	The photon-added coherent states $\ecoh$\cite{gsa} are defined as
\begin{equation}
\ecoh = {\com \coh \over{\sqrt{\langle\alpha\vert\ao^m \com\coh}}},
\end{equation}
where m is a nonegative integer.  The states $\ecoh$ exhibit nonclassical
features like phase squeezing and sub-Poissonian statistics. These
states are produced in the interaction of a two-level atom with a cavity
field initially prepared in the coherent state $\coh$\cite{gsa}.  In the
 present
 contribution it is shown that the photon-added coherent states can be 
interpreted as nonlinear coherent states.  This is done by showing that the
states $\ecoh$ obey the equation
\begin{equation}
\opfnm\ao\ecoh = \alpha \ecoh,
\end{equation}
with a suitable choice for the function $\opfnm$.  The operator $\opfnm$ is a 
valid operator for negative values of $m$ also.  The eigenstates of $\opfnm\ao$
with negative values of $m$ are constructed and their properties compared with 
those of $\ecoh$.

\section{CONSTRUCTION  OF $\opfnm$}

	In this section we construct the explicit form of the operator valued
function $\opfnm$. The coherent states $\coh$ satisfy, by definition,
\begin{equation}
\ao\coh = \alpha \coh,
\end{equation}
where $\alpha$ is a complex number.  Premultiplying both the sides of this
equation by $\com$ yields
\begin{equation}
\com \ao \coh = \alpha \com \coh,
\end{equation}
where $m$ is a nonnegative integer.
Using the commutation relation $[\ao, \co] = 1$, the above equation is written
as
\begin{equation}
(\ao \com - m \ao^{\dagger (m-1)})\coh = \alpha \com \coh,
\end{equation}
which, making use of the identity ${1\over{1+\co\ao}}\ao\co = 1$, leads to
\begin{equation}
(\ao - {m\over{1+\co\ao}}\ao)\com\coh = \alpha \com \coh.
\end{equation}
Comparing Eq. (7) and Eq. (3) gives the expression for $\opfnm$ as
\begin{equation}
\opfnm = 1-{m\over{1+\co\ao}}.
\end{equation}
This shows that the photon-added coherent states can be interpreted as
nonlinear coherent states.   These states are a class nonlinear
cohrent states than can be physically realized in the interaction of a two-
level atom with a cavity field that is initially prepared in the cohrent
state $\coh$.

\section{$\ecoh$ as deformed {\lowercase{m}}-photon number state $\vert
 \lowercase{m}\rangle$}

	In this section it is shown that the photon-added coherent states
can be written as the nonunitarily deformed number state .  This is achieved by 
the
method given by Shanta, et al \cite{shanta}.  Firstly, a brief review of the method is given.
Consider an "annihilation operator" $\hat{A}$ which annihilates a set 
of number states 
${\vert n_{i}\rangle,i=1,2,...k}$.  Then we can construct a sector $S_{i}$ by 
repeatedly applying $\hat{A}^\dagger$, the adjoint of $\hat{A}$, 
 on the number state $\vert n_{i}
\rangle$.  Thus we have $k$ sectors corresponding to the states
that are annihilated by $\hat{A}$.  A given sector may turnout to be either 
finite or infinite dimensional.  If a sector, say $S_{j}$, is of infinite 
dimension then we construct an operator $\hat{G_{j}}^\dagger$ such that the 
commutator $[\hat{A},\hat{G_{j}}^\dagger]=1$ holds in the sector.  Then the 
eigenstates of $\hat{A}$
can be written as $e^{\alpha\hat{G_{j}}^\dagger}\vert n_{j}\rangle$.  
If the operator $\hat{A}$ is of the form $\opfn\ao^p$, where $p$ is 
a nonnegative integer and $\opfn$ is a operator valued function of the 
number operator $\co\ao$, such that it annihilates the number state $\vert j
\rangle$ then $\hat{G_{j}}^\dagger$ is constructed as
\begin{equation}
\hat{G_{j}}^\dagger = {1\over p}{\hat{A}^\dagger}{1\over{\hat{A}\hat{A}^
\dagger}}{(\co\ao+p-j)}.
\end{equation}   

	The photon-added coherent states are the eigenstates of $\opfnm\ao$
with $\opfnm$ given by  Eq.(8).   This operator annihilates the vacuum state $\vert
0 \rangle$ and the $m$-photon state $\vert m \rangle$.  The
states in between the vacuum state and the m-photon state are not annihilated.
In this sense it is different from the m-photon annihilation operator $\ao^m$
which annihilates all the states $\vert i\rangle$, $i=0,1,2,..m$.
To discuss the case of $\ecoh$ 
let 
\begin{equation}
\hat{A}= (1-{m\over{1+\co\ao}})\ao.
\end{equation}
The adjoint $\hat{A}^\dagger$ is given by
\begin{equation}
\hat{A}^\dagger =\co (1-{m\over{1+\co\ao}}).
\end{equation}
We construct the sector $S_0$ by repeatedly applying $\hat{A}^\dagger$ on the
vacuum state $\vert 0 \rangle$.  $S_0$ is the set ${\vert i \rangle, i=0,1,2,...
m-1}$ and it is finite dimensional.  The secotr $S_m$, built by the repeated
application of $\hat{A}^\dagger$ on $\vert m\rangle$, is the set ${\vert i
\rangle, i=m,m+1,...}$ and it is of infinite dimension.  Hence we can construct
an operator $\hat{G}^\dagger$ such that $[\hat{A},\hat{G}^\dagger]=1$ holds in
$S_m$.
To construct  $\hat{G}^\dagger$, corresponding to the operator $\hat{A}$ given
by Eq.(10), we set $p=1$ and $j=m$ in Eq.(9) and this yields
\begin{equation}
\hat{G}^\dagger = \co.
\end{equation}
Thus on the sector $S_m$ we have $[\hat{A},\co ]=1$.   
 Thus the photon-added coherent states, which are the eigenstates of
$\hat{A}$, can be written as  $e^{\alpha\co}\vert m\rangle$.  However, this
is not a unitary deformation.  

\section{Eigenstates of $\opfnm\ao$ with negative $\lowercase{m}$}

	The form of $\opfnm$, given by Eq. (8), suggests that it is a well
 defined 
operator valued function, on the harmonic oscillator Hilbert space, for
negative integer values of $m$ also.  In this
section the nonlinear coherent states, with negative $m$ in the expression for
$\opfnm$, are constructed.  Denoting the eigenstates by $\necoh$, the equation
to determine them is
\begin{equation}
(1 + {m\over{1+\co\ao}})\ao\necoh = \alpha\necoh.
\end{equation}
Expanding $\necoh$ in terms of the number states $\ns$ as
\begin{equation}
\necoh = \sum_{n=0}^{\infty}c_{n}\ns,
\end{equation}
where $c_n$'s are the expansion coefficients and  substituting the expansion 
in Eq. (13) leads to the recursion relation
\begin{equation}
c_n = {m! \sqrt{n!}\alpha^n\over{(n+m)!}} c_0.
\end{equation}
The constant $c_0$ is determined by normalization.  The normalized $\necoh$
is given by
\begin{eqnarray}
\necoh = N \sum_{n=0}^{\infty} {\alpha^n\over{\sqrt{n!}(n+1)(n+2)...(n+m)}}
\ns;\\
 N^{-1}=\sqrt{\sum_{n=0}^{\infty}{\vert\alpha\vert^{2n}n!\over{(n+m)!^2}}}
={1\over{m!}}\sqrt{_2 F_2(1,1,m+1,m+1,\vert\alpha\vert^2)},\nonumber
\end{eqnarray}
where $_2 F_2(1,1,m+1,m+1,\vert\alpha\vert^2)$ is the generalized 
Hypergeometric function\cite{hyper}.  The number state expansion for the state
$\ecoh$ is\cite{gsa}
\begin{equation}
\ecoh = {\exp(-\vert\alpha\vert^2/2)\over{\sqrt{L_m(-\vert\alpha\vert^2)m!}}}
\sum_{n=0}^{\infty}{\alpha^n\sqrt{(m+n)!}\over{n!}}\vert n+m\rangle,
\end{equation}
where $L_m(x)$ is the Lauguerre polynomial of order m defined by
\cite{hyper}
\begin{equation}
L_m(x) = \sum_{n=0}^m{(-x)^nm!\over{(n!)^2(m-n)!}}.
\end{equation}

	The state $\necoh$ involves a superpostion of all the Fock states 
starting with the vacuum state $\vert 0\rangle$.  But in the state $\ecoh$
the Fock states $\vert 0\rangle, \vert 1\rangle ... \vert m-1\rangle$ are
not present.  This important difference leads to different limiting cases
of the states $\ecoh$ and $\necoh$ when $\alpha\rightarrow 0$.  In the limit
$\alpha\rightarrow0$ the state $\necoh$ becomes  the vacuum state $\vert 0
\rangle$, which has properties like a classical radiation field, irrespective
 of the value of $m$ and the state $\ecoh$ becomes the
Fock state $\vert m\rangle$ which is nonclassical.   
In the limit $m\rightarrow0$
the states $\ecoh$ and $\necoh$ become the coherent state $\coh$. 
 Thus, $\necoh$($\ecoh$) is a 
state that is intermediate between the vacuum state ( the number state $\vert 
m\rangle$) and the coherent state.

	The photon-added coherent states are obtained by the action of $\com$ 
on the coherent state.  The states $\necoh$ can be written in a 
similar form  using the inverse operators ${\iao}$ and $\ico$ \cite{clme}.
  These operators
are defined in terms of their actions on the number states $\ns$ as follows:
\begin{eqnarray}
\iao\ns & =&  {1\over{\sqrt{n+1}}} \vert {n+1} \rangle,\\
\ico\ns & =&{1\over{\sqrt{n}}} \vert {n-1} \rangle \quad
 \hbox {for n $\neq 0$} ,\\
\ico\vert 0\rangle & =& 0 .
\end{eqnarray}
Using these inverse operators and Eqn.(16) the state $\necoh$ can be written as
\begin{equation}
\necoh = N \icom\iaom\coh.
\end{equation}

	The states $\necoh$ correspond to the nonlinear coherent states with 
-m replacing m in $\opfnm$. However, they are obtained by the action of
$\icom\iaom$ on $\coh$ and not $\icom$ on $\coh$.   Using the method reviewed
in Section III we can show that
\begin{equation}
\necoh = e^{\alpha\hat{G}^\dagger}\vert 0\rangle,
\end{equation}
where $\hat{G}^\dagger = \co {{1+\co\ao}\over{1+m+\co\ao}}$. 

\section{SQUEEZING IN $\necoh$}

	The state $\necoh$ exhibits squeezing in both x- and p-quadratures.
The x- and p-quadratures are  given interms of $\ao$ and $\co$ by
\begin{equation}
x={{\ao+\co}\over{\sqrt{2}}},\quad  p={{\ao-\co}\over{i\sqrt{2}}}.
\end{equation}
The mean values of the relevant operators in the state $\necoh$ are 
\begin{eqnarray}
\langle\ao\rangle& =& \alpha N^2 \sum_{n=0}^{\infty}{\vert\alpha\vert^{2n}n!
\over{(n+m)!^2}}{(n+1)\over{(n+m+1)}},\\
\langle\ao^2\rangle& = &\alpha^2 N^2 \sum_{n=0}^{\infty}{\vert\alpha\vert^{2n}
n!\over{(n+m)!^2}}{(n+1)(n+2)\over{(n+m+1)(n+m+2)}},\\
\hbox{and}\nonumber\\
\langle\co\ao\rangle &= &N^2 \sum_{n=0}^{\infty}{\vert\alpha\vert^{2n}n!
\over{(n+m)!^2}}n.
\end{eqnarray}
The mean values of $\co$ and $\ao^{\dagger 2}$ are obtained by taking the 
complex conjugates of $\langle\ao\rangle$ and $\langle\ao^2\rangle$
respectively. The uncertainty in x is 
\begin{equation}
\langle x^2\rangle-\langle x\rangle^2 = {1\over{2}}[1+2\langle \co\ao\rangle
+\langle\ao^2\rangle+\langle\cot\rangle-\langle\ao\rangle^2-\langle\co
\rangle^2-2\langle\ao\rangle\langle\co\rangle],
\end{equation}
and that in p is
\begin{equation}
\langle p^2\rangle-\langle p\rangle^2 = {1\over{2}}[1+2\langle \co\ao\rangle
-\langle\ao^2\rangle-\langle\cot\rangle+\langle\ao\rangle^2+\langle\co
\rangle^2-2\langle\ao\rangle\langle\co\rangle].
\end{equation}

	In Fig. 1 the variance in p is shown for real  $\alpha$ for various
values of m.  As expected the uncertainty in p is close to $1\over{2}$, the 
unertainty in p in the 
vacuum state, when $\alpha$ is close to zero. In the case of the state $\ecoh$
the variance is close to $m+{1\over{2}}$ when $\alpha$ is close to zero.  For
real values of $\alpha$ the p-quadrature is always squeezed, $\langle p^2
\rangle -\langle p\rangle^2 < {1\over{2}}$,  for the state
$\necoh$. For large values of $\alpha$ the variance in p approaches that of
the coherent state.  If $\alpha$ becomes $i\alpha$ the expression for variance
in p becomes the expression for variance in x.  Since p shows squeezing for
real $\alpha$ the x-quadrature exhibits squeezing for imaginary $\alpha$. 

\section{PHOTON STATISTICS OF $\necoh$}
	The photon number distribution $p(n)$ for the state $\necoh$ is
\begin{eqnarray}
p(n) &=& \vert \langle n \necoh\vert^2,\nonumber\\
&=& N {\vert\alpha\vert^{2n}n!\over{(n+m)!^2}}.
\end{eqnarray}
When $m=0$ the distribution becomes the Poissonian distribution whose mean
value is $\vert\alpha\vert^2$.  

	A measure of the variance of the photon number distribution is given
 by the Mandel's q parameter\cite{mandl},
\begin{equation}
q=\frac{\langle(\triangle\hat{n})^2\rangle - \langle\hat{n}\rangle}{\langle
\hat{n}\rangle}.
\end{equation}
The coherent states have q as zero. Negative value of q indicates that
the distribution $p(n)$ is sub-Poissonian and it is a nonclassical feature.
The photon-added coherent states $\ecoh$ are always sub-Poissonian for all
values of m.  For the state $\necoh$ the mean values of $\hat{n}$ and 
$\hat{n}^2$ are given by 
\begin{eqnarray}
\langle \hat{n}\rangle &=& N^2 \sum_{n=0}^{\infty}{\vert\alpha\vert^{2n}n!
\over{(n+m)!^2}}n,\\
\langle\hat{n}^2\rangle &=& N^2 \sum_{n=0}^{\infty}{\vert\alpha\vert^{2n}n!
\over{(n+m)!^2}}n^2.
\end{eqnarray}

	In Fig. 2 the q-parameter, calculated from Eqs. (31)-(33), for the
 state $\necoh$ is shown as a function of $\vert\alpha\vert$.  The states
 $\necoh$ have q always greater than zero indicating that they are
 super-Poissonian.   For small values of $\alpha$ the q-parameter is close
to zero and for large  values of $\alpha$ it approaches zero.

\section{ SUMMARY}
	The photon-added coherent states are nonlinear coherent states.  They
are the eigenstates of the operator $(1 - {m\over{1+\co\ao}})\ao$ when $m$ takes
positive integer values.  This operator is a meaningful operator when $m$ takes
negative integer values also.  The corresponding eigenstates $\necoh$ are
nonclassical.  The photon-added coherent state $\ecoh$ results from the
action of $\com$ on the coherent state $\coh$ while the state $\necoh$
comes from the action of $\icom\iaom$ on the coherent
 state $\coh$.
Both $\ecoh$ and $\necoh$ show squeezing.  While $\ecoh$ is
sub-Poissonian the state $\necoh$ is not.  The states $\ecoh$ and $\necoh$
become $\vert m \rangle$ and $\vert 0 \rangle$ respectively in the limit 
$\alpha\rightarrow 0$ but they become the coherent state $\coh$
when $m\rightarrow 0$.  The states $\ecoh$ and $\necoh$  are the result of
deforming the number states  $\vert m\rangle$ and $\vert 0 \rangle$
respectively. 

	The author acknowledges Drs. V. Balakrishnan,  M. V. Satyanarayana
and D. Sahoo for useful discussions.

\begin{figure}

\caption{Uncertainty S, $\langle p^2\rangle -\langle p\rangle^2$,  in p as a
 function of $\alpha$ for m=1, m=5 and m=10 
for the state $\necoh$.
 The real $\alpha$ is denoted as $r$.}
\label{fig1}
\end{figure}
\begin{figure}
\caption{Mandel's q parameter as a function of $\vert\alpha\vert$ for 
  m = 1, m = 5 and m = 10.  $\vert\alpha\vert$ is denoted as $r$}
\label{fig2}

\end{figure}

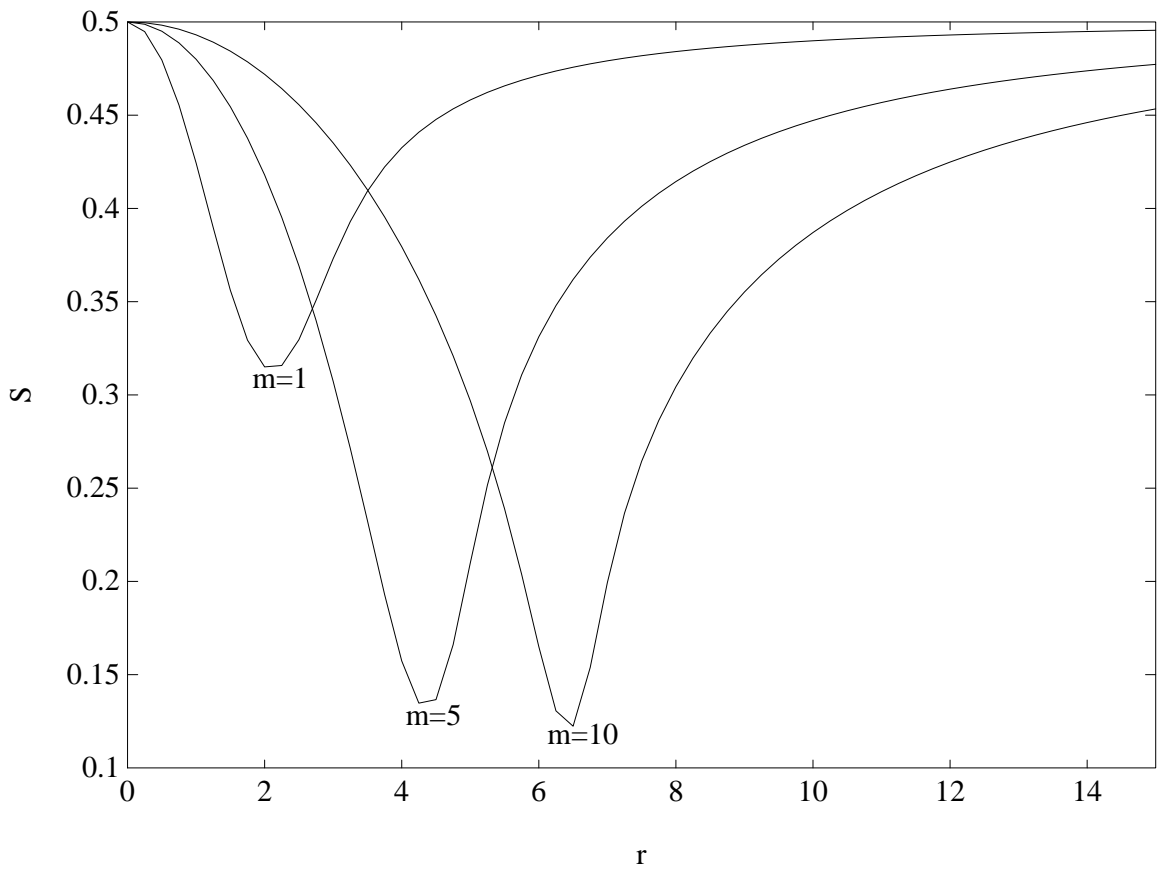
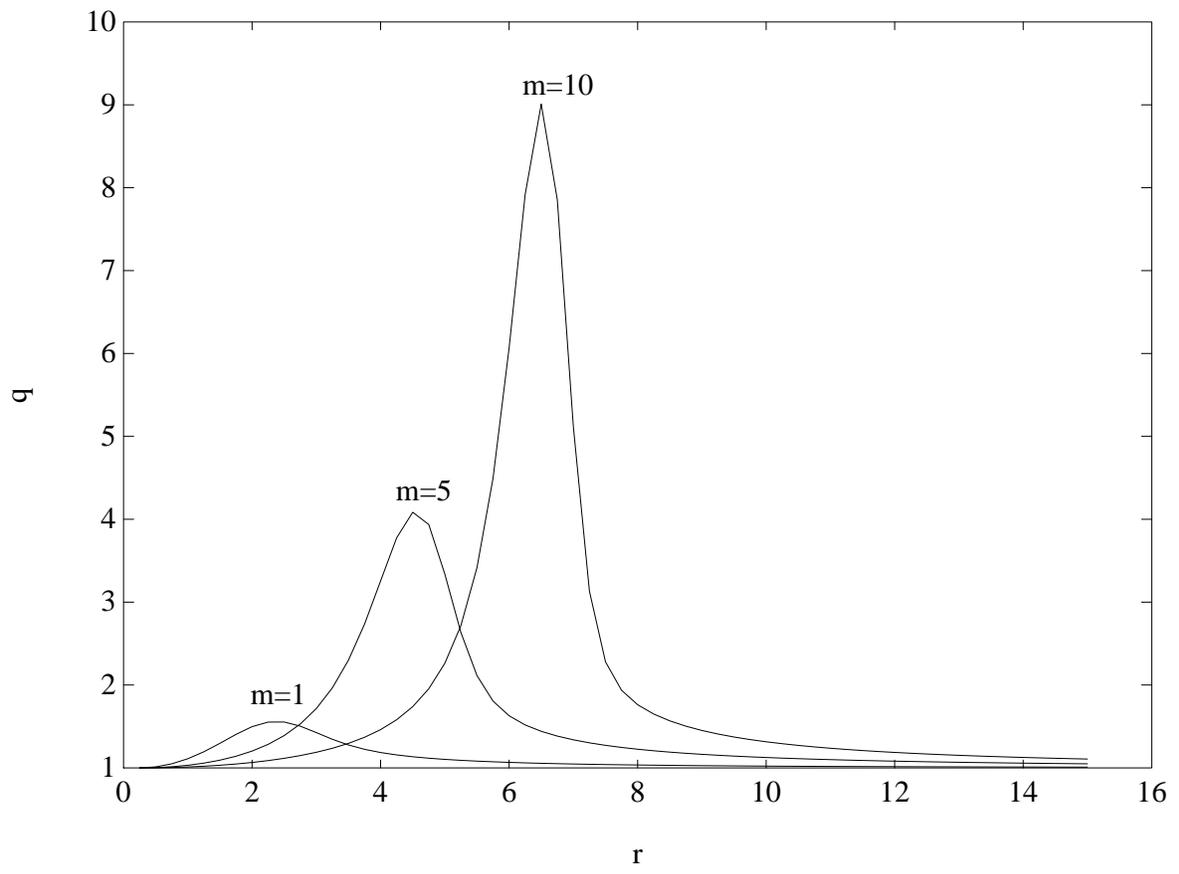
\end{document}